\documentclass[preprint,showpacs,showkeys,preprintnumbers,amsmath,amssymb]{revtex4}

\usepackage{bm}

\begin{document}

\title{Approximate symmetry reduction approach: infinite series reductions to the KdV-Burgers equation}

\author{Xiaoyu Jiao$^{1}$, Ruoxia Yao$^{1,2,3}$, Shunli Zhang$^4$ and S. Y.
Lou$^{1,2}$}

\affiliation{$^1$Department of Physics, Shanghai Jiao Tong
University, Shanghai, 200240,  China\\
$^2$Department of Physics, Ningbo University, Ningbo, 315211,  China\\
$^3$School of Computer Science, Shaanxi Normal University, Xi'an,
710062, China\\
$^4$Department of Mathematics, Northwest University, Xi'an, 710069,
China}

\begin{abstract}
For weak dispersion and weak dissipation cases, the
(1+1)-dimensional KdV-Burgers equation is investigated in terms of
approximate symmetry reduction approach. The formal coherence of
similarity reduction solutions and similarity reduction equations of
different orders enables series reduction solutions. For weak
dissipation case, zero-order similarity solutions satisfy the
Painlev\'e II, Painlev\'e I and Jacobi elliptic function equations.
For weak dispersion case, zero-order similarity solutions are in the
form of Kummer, Airy and hyperbolic tangent functions. Higher order
similarity solutions can be obtained by solving linear ordinary
differential equations.
\end{abstract}


\maketitle

\section{Introduction}
Nonlinear problems arise in many fields of science and engineering.
Lie group theory \cite{Olver,Bluman1,Bluman2} greatly simplifies
many nonlinear partial differential equations. Exact analytical
solutions are nonetheless difficult to study in general.
Perturbation theory \cite{Cole,Dyke,Nayfeh} was thus developed and
it plays an essential role in nonlinear science, especially in
finding approximate analytical solutions to perturbed partial
differential equations.

The integration of Lie group theory and perturbation theory yields
two distinct approximate symmetry reduction methods. The first
method due to Baikov, et al \cite{Baikov1,Baikov2} generalizes
symmetry group generators to perturbation forms. For the second
method proposed by Fushchich, et al \cite{Fushchich}, dependent
variables are expanded in perturbation series and approximate
symmetry of the original equation is decomposed into exact symmetry
of the system resulted from perturbation. The second method is
superior to the first one from the comparison in Refs.
\cite{Pakdemirli,Wiltshire}.

The well known Korteweg-de Vries-Burgers (KdV-Burgers) equation
\begin{equation}
u_t+6uu_x+\mu u_{xxx}+\nu u_{xx}=0,\label{kdvburgers}
\end{equation}
with $\mu$ and $\nu$ constant coefficients, is widely used in the
many physical fields especially in fluid dynamics. The effects of
nonlinearity ($6uu_{x}$), dispersion ($\mu u_{xxx}$) and dissipation
($\nu u_{xx}$) are incorporated in this equation which simulates the
propagation of waves on an elastic tube filled with a viscous fluid
\cite{Johnson}, and the flow of liquids containing gas bubbles
\cite{Wijngaarden} and turbulence \cite{Gao,Liu}, etc.

Johnson \cite{Johnson} inspected the travelling wave solutions to
the weak dissipation ($\nu\ll1$) KdV-Burgers equation
(\ref{kdvburgers}) in the phase plane by a perturbation method and
developed formal asymptotic expansion for the solution. Tanh
function method was applied to Eq. (\ref{kdvburgers}) in the limit
of weak dispersion ($\mu\ll1$) in a perturbative way
\cite{Malfliet}. In Refs. \cite{Marchant,Allen}, perturbation
analysis was also applied to the perturbed KdV equations
\begin{equation}
\eta_t+6\eta\eta_x+\eta_{xxx}+\alpha c_1\eta_{xx}+\alpha
c_2\eta_{xxxx}+\alpha c_3(\eta\eta_x)_x=0,\ \alpha\ll1,
\end{equation}
and
\begin{equation}
u_t+uu_x+u_{xxx}=\epsilon\alpha u+\epsilon\beta u_{xx},\
\epsilon\ll1,
\end{equation}
respectively.

Eq. (\ref{kdvburgers}) can also be manipulated by means of
approximate symmetry reduction approach. Section II and section III
are devoted to applying approximate symmetry reduction approach to
Eq. (\ref{kdvburgers}) under the case of weak dissipation
($\nu\ll1$) and weak dispersion ($\mu\ll1$), respectively. Section
IV is conclusion and discussion of the results.
\section{Approximate symmetry reduction approach to weak dissipation KdV-Burgers equation}
According to the perturbation theory, solutions to perturbed partial
differential equations can be expressed as a series containing a
small parameter. Specifically, we suppose that the weak dissipation
($\nu\ll1$) KdV-Burgers equation (\ref{kdvburgers}) has the solution
\begin{equation}\label{seriessol1}
u=\sum_{k=0}^\infty\nu^ku_k,
\end{equation}
where $u_k$ are functions of $x$ and $t$, and solve the following
system
\begin{equation}\label{approximateeq1}
u_{k,t}+6\sum_{i=0}^ku_{k-i}u_{i,x}+\mu u_{k,xxx}+u_{k-1,xx}=0,\
(k=0,\ 1,\ \cdots)
\end{equation}
with $u_{-1}=0$, which is obtained by inserting Eq.
(\ref{seriessol1}) into Eq. (\ref{kdvburgers}) and vanishing the
coefficients of different powers of $\nu$.

The next crucial step is to study symmetry reduction of the above
system via the Lie symmetry approach \cite{Lou}. To that end, we
construct the Lie point symmetries
\begin{equation}\label{liesym}
\sigma_k=Xu_{kx}+Tu_{kt}-U_k,\ (k=0,\ 1,\ \cdots),
\end{equation}
where $X$, $T$ and $U_k$ are functions with respect to $x$, $t$ and
$u_i,\ (i=0,\ 1,\ \cdots)$. The linearized equations for Eq.
(\ref{approximateeq1}) are:
\begin{equation}\label{linearizedeq1}
\sigma_{k,t}+6\sum_{i=0}^k\left(\sigma_{k-i}u_{i,x}+u_{k-i}\sigma_{i,x}\right)+\mu
\sigma_{k,xxx}+\sigma_{k-1,xx}=0,\ (k=0,\ 1,\ \cdots)
\end{equation}
with $\sigma_{-1}=0$. Eq. (\ref{linearizedeq1}) means that Eq.
(\ref{approximateeq1}) is invariant under the transformations
$u_k\rightarrow u_k+\varepsilon\sigma_k\ (k=0,\ 1,\ \cdots)$ with an
infinitesimal parameter $\varepsilon$.

There are infinite number of equations in Eqs.
(\ref{approximateeq1}) and (\ref{linearizedeq1}) and infinite number
of arguments in $X$, $T$ and $U_k\ (k=0,\ 1,\ \cdots)$. To simplify
the problem, we begin the discussion from finite number of
equations.

Confining the range of $k$ to $\{k|k=0,\ 1,\ 2\}$ in Eqs.
(\ref{approximateeq1}), (\ref{liesym}) and (\ref{linearizedeq1}), we
see that $X$, $T$, $U_0$, $U_1$ and $U_2$ are functions with respect
to $x$, $t$, $u_0$, $u_1$ and $u_2$. In this case, the determining
equations can be obtained by substituting Eq. (\ref{liesym}) into
Eq. (\ref{linearizedeq1}), eliminating $u_{0,t}$, $u_{1,t}$ and
$u_{2,t}$ in terms of Eq. (\ref{approximateeq1}) and vanishing all
coefficients of different partial derivatives of $u_0$, $u_1$ and
$u_2$. Some of the determining equations read
\begin{displaymath}
T_x=T_{u_0}=T_{u_1}=T_{u_2}=0,
\end{displaymath}
from which we have $T=T(t)$. Considering this condition, we choose
the simplest equations for $X$
\begin{displaymath}
X_{u_0}=X_{u_1}=X_{u_2}=0,
\end{displaymath}
from which we have $X=X(x,t)$. Considering this condition, we choose
the simplest equations for $U_0$, $U_1$ and $U_2$
\begin{displaymath}
U_{0,xu_2}=U_{0,u_0u_0}=U_{0,u_0u_1}=U_{0,u_0u_2}=U_{0,u_1u_1}=U_{0,u_1u_2}=U_{0,u_2u_2}=0,
\end{displaymath}
\begin{displaymath}
U_{1,u_0u_0}=U_{1,u_0u_1}=U_{1,u_0u_2}=U_{1,u_1u_1}=U_{1,u_1u_2}=U_{1,u_2u_2}=0,
\end{displaymath}
\begin{displaymath}
U_{2,u_0u_0}=U_{2,u_0u_1}=U_{2,u_0u_2}=U_{2,u_1u_1}=U_{2,u_1u_2}=U_{2,u_2u_2}=0,
\end{displaymath}
of which the solutions are
\begin{displaymath}
U_0=F_1(x,t)u_0+F_2(x,t)u_1+F_3(t)u_2+F_4(x,t),
\end{displaymath}
\begin{displaymath}
U_1=F_5(x,t)u_0+F_6(x,t)u_1+F_7(x,t)u_2+F_8(x,t),
\end{displaymath}
\begin{displaymath}
U_2=F_9(x,t)u_0+F_{10}(x,t)u_1+F_{11}(x,t)u_2+F_{12}(x,t).
\end{displaymath}
Under these relations, the determining equations are simplified to
\begin{displaymath}
X_{xx}=F_2=F_3=F_5=F_7=F_8=F_9=F_{10}=F_{12}=F_{1,x}=F_{4,xx}=F_{4,t}=F_{6,x}=F_{11,x}=0,
\end{displaymath}
\begin{displaymath}
T_t=3X_x,\ X_t=6F_4,\ F_{1,t}=-6F_{4,x},\ F_{6,t}=-6F_{4,x},\
F_{11,t}=-6F_{4,x},
\end{displaymath}
\begin{displaymath}
T_t=X_x-F_1,\ T_t=2X_x-F_1+F_6,\ T_t=X_x+F_{11}-2F_6,\
T_t=2X_x-F_6+F_{11},
\end{displaymath}
which provide us with
\begin{displaymath}
X=6at+cx+x_0,\ T=3ct+t_0,\ U_0=-2cu_0+a,\ U_1=-cu_1,\ U_2=0,
\end{displaymath}
where $a$, $c$, $x_0$ and $t_0$ are arbitrary constants.

Similarly, limiting the range of $k$ to $\{k|k=0,\ 1,\ 2,\ 3\}$ in
Eqs. (\ref{approximateeq1}), (\ref{liesym}) and
(\ref{linearizedeq1}) where $X$, $T$, $U_0$, $U_1$, $U_2$ and $U_3$
are functions with respect to $x$, $t$, $u_0$, $u_1$, $u_2$ and
$u_3$, we repeat the calculation as before and obtain
\begin{displaymath}
X=6at+cx+x_0,\ T=3ct+t_0,\ U_0=-2cu_0+a,\ U_1=-cu_1,\ U_2=0,\
U_3=cu_3,
\end{displaymath}
where $a$, $c$, $x_0$ and $t_0$ are arbitrary constants.

With more similar calculation considered, we see that $X$, $T$ and
$U_k\ (k=0,\ 1,\ \cdots)$ are formally coherent, i.e.,
\begin{equation}\label{determiningeqsol1}
X=6at+cx+x_0,\ T=3ct+t_0,\ U_k=(k-2)cu_k+a\delta_{k,0},\ (k=0,\ 1,\
\cdots)
\end{equation}
where $a$, $c$, $x_0$ and $t_0$ are arbitrary constants. The
notation $\delta_{k,0}$ satisfying $\delta_{0,0}=1$ and
$\delta_{k,0}=0\ (k\neq0)$ is adopted in the following text.
Subsequently, solving the characteristic equations
\begin{equation}\label{characteristiceq}
\frac{{\rm d}x}{X}=\frac{{\rm d}t}{T},\ \frac{{\rm d}u_0}{U_0}
=\frac{{\rm d}t}{T},\ \frac{{\rm d}u_1}{U_1}=\frac{{\rm d}t}{T},\
\cdots,\ \frac{{\rm d}u_k}{U_k}=\frac{{\rm d}t}{T},\ \cdots
\end{equation}
leads to the similarity solutions to Eq. (\ref{approximateeq1})
which can be distinguished in the following two subcases.
\subsection{Symmetry reduction of the Painlev\'e II solutions}
When $c\neq 0$, for brevity of the results, we rewrite the constants
$a$, $x_0$ and $t_0$ as $ca$, $cx_0$ and $ct_0$, respectively.
Solving $\frac{{\rm d}x}{X}=\frac{{\rm d}t}{T}$ in Eq.
(\ref{characteristiceq}) leads to the invariant
\begin{equation}
I(x,t)=\xi=(x-3at+x_0-3at_0)(3t+t_0)^{-\frac{1}{3}}.
\end{equation}
In the same way, we get other invariants
\begin{equation}
I_0(x,t,u_0)=P_0=\frac{1}{2}(3t+t_0)^\frac{2}{3}(2u_0-a)
\end{equation}
and
\begin{equation}
I_k(x,t,u_k)=P_k=u_k(3t+t_0)^{-\frac{1}{3}(k-2)},\ (k=1,\ 2,\
\cdots)
\end{equation} from $\frac{{\rm d}u_0}{U_0}
=\frac{{\rm d}t}{T}$ and $\frac{{\rm d}u_k}{U_k}=\frac{{\rm d}t}{T}\
(k=1,\ 2,\ \cdots)$ respectively. Viewing $P_k\ (k=0,\ 1,\ \cdots)$
as functions of $\xi$, we get the similarity solutions to Eq.
(\ref{approximateeq1})
\begin{equation}\label{kdvburgers1similaritysol1}
u_k=(3t+t_0)^{\frac{1}{3}(k-2)}P_k(\xi)+\frac{1}{2}a\delta_{k,0},\
(k=0,\ 1,\ \cdots)
\end{equation}
with the similarity variable
\begin{equation}
\xi=(x-3at+x_0-3at_0)(3t+t_0)^{-\frac{1}{3}}.
\end{equation}

Accordingly, the series reduction solution to Eq. (\ref{kdvburgers})
is derived
\begin{equation}\label{kdvburgers1sol1}
u=\frac{a}{2}+\sum_{k=0}^\infty\nu^k(3t+t_0)^{\frac{1}{3}(k-2)}P_k(\xi),
\end{equation}
and the related similarity reduction equations are
\begin{equation}\label{kdvburgers1reductioneq1}
\mu P_{k,\xi\xi\xi}+6\sum_{i=0}^kP_{k-i}P_{i,\xi}-\xi
P_{k,\xi}+(k-2)P_k+P_{k-1,\xi\xi}=0,\ (k=0,\ 1,\ \cdots)
\end{equation}
with $P_{-1}=0$. When $k=0$, Eq. (\ref{kdvburgers1reductioneq1}) is
equivalent to the Painlev\'e II type equation. The $n$th $(n>0)$
similarity reduction equation is actually a third order
\textit{linear} ordinary differential equation of $P_n$ when the
previous $P_0$, $P_1$, $\cdots$, $P_{n-1}$ are known, since Eq.
(\ref{kdvburgers1reductioneq1}) is just
$$
\mu P_{k,\xi\xi\xi}+6(P_0P_{k,\xi}+P_kP_{0,\xi})-\xi
P_{k,\xi}+(k-2)P_k=f_k(\xi),\ (k=1,\ 2,\ \cdots)\eqno(16')
$$
where $f_k$ is a only function of $\{P_0,\ P_1,\ \cdots,\ P_{k-1}\}$
\begin{displaymath}
f_k(\xi)=-6\sum_{i=1}^{k-1}P_{k-i}P_{i,\xi}-P_{k-1,\xi\xi}.
\end{displaymath}
\subsection{Symmetry reduction of the Painlev\'e I solutions}
When $c=0$ and $t_0\neq0$, we rewrite $a$ and $x_0$ as $at_0$ and
$x_0t_0$, respectively. The similarity solutions are
\begin{equation}\label{kdvburgers1similaritysol2}
u_k=(at+\frac{x_0}{6})\delta_{k,0}+P_k(\xi),\ (k=0,\ 1,\ \cdots)
\end{equation}
with the similarity variable $\xi=-x+3at^2+x_0t$. Hence, the series
reduction solution to Eq. (\ref{kdvburgers}) is
\begin{equation}\label{kdvburgers1sol2}
u=at+\frac{x_0}{6}+\sum_{k=0}^\infty\nu^kP_k(\xi),
\end{equation}
in which $P_k(\xi)$ satisfys
\begin{equation}\label{kdvburgers1reductioneq2}
\mu
P_{k,\xi\xi}+3\sum_{i=0}^kP_{k-i}P_i-P_{k-1,\xi}-a\delta_{k,0}\xi+A_k=0,\
(k=0,\ 1,\ \cdots)
\end{equation}
where $P_{-1}=0$ and $A_k$ are arbitrary integral constants.

When $k=0$, Eq. (\ref{kdvburgers1reductioneq2}) is equivalent to the
Painlev\'e I type equation provided that $a\neq0$.

When $k=0$, $a=0$ and $A_0=\frac{4}{3}\mu^2p_1^4(m^2-1-m^4)$, Eq.
(\ref{kdvburgers1reductioneq2}) can be solved by the Jacobi elliptic
function,
\begin{equation}\label{cn}
P_0=\frac{2}{3}\mu p_1^2(1-2m^2)+2\mu
m^2p_1^2\mathrm{cn}^2(p_1\xi+p_2,m),
\end{equation}
where $p_1$, $p_2$ and $m$ are arbitrary constants.

When $k>0$, an equivalent form of Eq.
(\ref{kdvburgers1reductioneq2}) is
$$
\mu P_{k,\xi\xi}+6P_kP_0=g_k(\xi),\ (k=1,\ 2,\ \cdots)\eqno(19')
$$
where $g_k(\xi)$ is a function of $\{P_0,\ P_1,\ \cdots,\ P_{k-1}\}$
as follows
\begin{displaymath}
g_k(\xi)=-3\sum_{i=1}^{k-1}P_{k-i}P_i+P_{k-1,\xi}-A_k.
\end{displaymath}
From Eq. $(19')$, we see that Eq. \eqref{kdvburgers1reductioneq2} is
a second order linear ordinary differential equation of $P_k$ and
can be integrated out step by step when $a=0$. The results read
\begin{equation}
P_k=P_{0,\xi}[C_k+\mu^{-1}\int P_{0,\xi}^{-2}(B_k+\int
P_{0,\xi}g_k\mathrm{d}\xi)\mathrm{d}\xi],
\end{equation}
with arbitrary integral constants $B_k$ and $C_k$.
\section{Approximate symmetry reduction approach to weak dispersion KdV-Burgers equation}
We search for series reduction solutions to weak dispersion
$(\mu\ll1)$ KdV-Burgers equation (\ref{kdvburgers}). The process is
similar to the section II. A system of partial differential
equations
\begin{equation}\label{approximateeq2}
u_{k,t}+6\sum_{i=0}^ku_{k-i}u_{i,x}+\nu u_{k,xx}+u_{k-1,xxx}=0,\
(k=0,\ 1,\ \cdots)
\end{equation}
with $u_{-1}=0$, is obtained by plugging the perturbation series
solution
\begin{equation}\label{seriessol2}
u=\sum_{k=0}^\infty\mu^ku_k,
\end{equation}
into Eq. (\ref{kdvburgers}) and vanishing the coefficients of
different powers of $\mu$.

It is easily seen that the linearized equations related to Eq.
\eqref{approximateeq2} are
\begin{equation}\label{linearizedeq2}
\sigma_{k,t}+6\sum_{i=0}^k(\sigma_{k-i}u_{i,x}+u_{k-i}\sigma_{i,x})+\nu
\sigma_{k,xx}+\sigma_{k-1,xxx}=0,\ (j=0,\ 1,\ \cdots)
\end{equation}
with $\sigma_{-1}=0$.

The Lie point symmetries (\ref{liesym}) satisfy the linearized
equations \eqref{linearizedeq2} under the approximate equations
(\ref{approximateeq2}). Restricting the range of $k$ to $\{k|k=0,\
1,\ 2\}$ in Eqs. (\ref{liesym}), (\ref{approximateeq2}) and
(\ref{linearizedeq2}), we see that $X$, $T$, $U_0$, $U_1$ and $U_2$
are functions with respect to $x$, $t$, $u_0$, $u_1$ and $u_2$. The
determining equations are derived by substituting Eq. (\ref{liesym})
into Eq. (\ref{linearizedeq2}), eliminating $u_{0,t}$, $u_{1,t}$ and
$u_{2,t}$ in terms of Eq. (\ref{approximateeq2}) and vanishing
coefficients of different partial derivatives of $u_0$, $u_1$ and
$u_2$. Some of the determining equations are
\begin{displaymath}
T_x=T_{u_0}=T_{u_1}=T_{u_2}=0,
\end{displaymath}
from which we get $T=T(t)$. Considering this condition, the simplest
equations for $X$ in the determining equations are
\begin{displaymath}
X_{u_0}=X_{u_1}=X_{u_2}=0,
\end{displaymath}
from which we get $X=X(x,t)$. Considering this condition, we select
the simplest equations for $U_0$ and $U_1$
\begin{displaymath}
U_{0,u_0u_0}=U_{0,a_1}=U_{0,a_2}=0,
\end{displaymath}
\begin{displaymath}
U_{1,xu_0}=U_{1,u_0u_0}=U_{1,u_0u_1}=U_{1,u_1u_1}=U_{1,u_2}=0,
\end{displaymath}
with the solution $U_0=F_1(x,t)u_0+F_2(x,t)$ and
$U_1=F_3(t)u_0+F_4(x,t)u_1+F_5(x,t)$. From the reduced determining
equations, the simplest equations for $U_2$ read
\begin{displaymath}
U_{2,u_0u_0}=U_{2,u_0u_1}=U_{2,u_0u_2}=U_{2,u_1u_1}=U_{2,u_1u_2}=U_{2,u_2u_2}=0,
\end{displaymath}
leading to $U_2=F_6(x,t)u_0+F_7(x,t)u_1+F_8(x,t)u_2+F_9(x,t)$.

Combined with these conditions, the determining equations are
simplified to
\begin{displaymath}
X_{xx}=F_3=F_5=F_6=F_7=F_9=F_{1,x}=F_{2,xxx}=F_{4,x}=F_{8,x}=0,
\end{displaymath}
\begin{displaymath}
T_t=2X_x,\ X_t=6F_2,\ F_{1,t}=-6F_{2,x},\ F_{2,t}=-\nu F_{2,xx},\
F_{4,t}=-6F_{2,x},\ F_{8,t}=-6F_{2,x},
\end{displaymath}
\begin{displaymath}
T_t=X_x-F_1,\ T_t=X_x-2F_4+F_8,\ T_t=3X_x-F_1+F_4,\
T_t=3X_x-F_4+F_8,
\end{displaymath}
It is easily seen that
\begin{displaymath}
X=6at+cx+x_0,\ T=2ct+t_0,\ U_0=-cu_0+a,\ U_1=-2cu_1,\ U_2=-3cu_2,
\end{displaymath}
where $a$, $c$, $x_0$ and $t_0$ are arbitrary constants.

In like manner, we obtain
\begin{displaymath}
X=6at+cx+x_0,\ T=2ct+t_0,\ U_0=-cu_0+a,\ U_1=-2cu_1,\ U_2=-3cu_2,\
U_3=-4cu_3,
\end{displaymath}
where $a$, $c$, $x_0$ and $t_0$ are arbitrary constants.

Repeating similar calculation several times, we summarize the
solutions to the determining equations
\begin{equation}\label{determiningeqsol2}
X=6at+cx+x_0,\ T=2ct+t_0,\ U_k=-(k+1)cu_k+a\delta_{k,0},\ (k=0,\ 1,\
\cdots)
\end{equation}
where $a$, $c$, $x_0$ and $t_0$ are arbitrary constants. The
similarity solutions to Eq. (\ref{approximateeq2}) from solving the
characteristic equations (\ref{characteristiceq}) are discussed in
the following two subcases.
\subsection{Symmetry reduction of the Kummer function solutions}
When $c\neq0$, for brevity of the results, we rewrite the constants
$a$, $x_0$ and $t_0$ as $ca$, $cx_0$ and $ct_0$, respectively.
Solving $\frac{{\rm d}x}{X}=\frac{{\rm d}t}{T}$ in Eq.
(\ref{characteristiceq}) results in the invariant
\begin{equation}
I(x,t)=\xi=(x-6at-6at_0+x_0)(2t+t_0)^{-\frac{1}{2}}.
\end{equation}
Likewise, we get other invariants
\begin{equation}
I_0(x,t,u_0)=P_0=(2t+t_0)^\frac{1}{2}(u_0-a)
\end{equation}
and
\begin{equation}
I_k(x,t,u_k)=P_k=u_k(2t+t_0)^{\frac{1}{2}(k+1)}\ (k=1,\ 2,\ \cdots)
\end{equation}
from $\frac{{\rm d}u_0}{U_0}=\frac{{\rm d}t}{T}$ and $\frac{{\rm
d}u_k}{U_k}=\frac{{\rm d}t}{T}\ (k=1,\ 2,\ \cdots)$ respectively.
Considering $P_k\ (k=0,\ 1,\ \cdots)$ as functions of $\xi$, we get
the similarity solutions to Eq. (\ref {approximateeq2})
\begin{equation}\label{kdvburgers2similaritysol1}
u_k=(2t+t_0)^{-\frac{1}{2}(k+1)}P_k(\xi)+a\delta_{k,0},\ (k=0,\ 1,\
\cdots)
\end{equation}
with the similarity variable
$\xi=(x-6at-6at_0+x_0)(2t+t_0)^{-\frac{1}{2}}$, and the series
reduction solution to Eq. (\ref{kdvburgers}) is
\begin{equation}\label{kdvburgers2sol1}
u=a+\sum_{k=0}^\infty\mu^k(2t+t_0)^{-\frac{1}{2}(k+1)}P_k(\xi),
\end{equation}
where $P_k(\xi)$ are subject to
\begin{equation}\label{kdvburgers2reductioneq1}
\nu P_{k,\xi\xi}+6\sum_{i=0}^kP_{k-i}P_{i,\xi}-\xi
P_{k,\xi}-(k+1)P_k+P_{k-1,\xi\xi\xi}=0,\ (k=0,\ 1,\ \cdots)
\end{equation}
with $P_{-1}=0$.

When $k=0$, Eq. (\ref{kdvburgers2reductioneq1}) has the Kummer
function solution
\begin{eqnarray}
P_0=\frac{(3C_1-1)\nu[3C_1C_2\mathrm{K}_1(\frac{3}{2}(1-C_1),\frac{3}{2},\frac{\xi^2}{2\nu})-
2\mathrm{K}_2(\frac{3}{2}(1-C_1),\frac{3}{2},\frac{\xi^2}{2\nu})]}{6\xi[C_2\mathrm{K}_1(\frac{1}{2}(1-3C_1),\frac{3}{2},
\frac{\xi^2}{2\nu})+\mathrm{K}_2(\frac{1}{2}(1-3C_1),\frac{3}{2},\frac{\xi^2}{2\nu})]}+\frac{C_1\nu}{\xi},
\end{eqnarray}
where $C_1$ and $C_2$ are arbitrary constants, and the two types of
Kummer functions ${\rm K_1}(p,\ q,\ z)$ and ${\rm K_2}(p,\ q,\ z)$
solve the differential equation
\begin{displaymath}
zy''(z)+(q-z)y'(z)-py(z)=0.
\end{displaymath}

When $(k>0)$, we rearrange the terms in Eq.
(\ref{kdvburgers2reductioneq1}) as
\begin{equation}\label{kdvburgers2reductioneq1a}
\nu P_{k,\xi\xi}+6(P_0P_{k,\xi}+P_kP_{0,\xi})-\xi
P_{k,\xi}-(k+1)P_k=-6\sum_{i=1}^{k-1}P_{k-i}P_{i,\xi}-P_{k-1,\xi\xi\xi},\
(k=1,\ 2,\ \cdots)
\end{equation}
which is a second order linear ordinary differential equation of
$P_k$ when the previous $P_0$, $P_1$, $\cdots$, $P_{k-1}$ are known.

\subsection{Symmetry reduction of Airy function and hyperbolic tangent function solutions}
When $c=0$ and $t_0\neq0$, we rewrite the constants $a$ and $x_0$ as
$at_0$ and $x_0t_0$, respectively. It is easily seen that the
similarity solutions are
\begin{equation}\label{kdvburgers2similaritysol2}
u_k=(at+\frac{x_0}{6})\delta_{k,0}+P_k(\xi),\ (k=0,\ 1,\ \cdots)
\end{equation}
with the similarity variable $\xi=-x+3at^2+x_0t$, and the series
reduction solution to Eq. (\ref{kdvburgers}) is
\begin{equation}
u=at+\frac{x_0}{6}+\sum_{k=0}^\infty\mu^kP_k(\xi),
\end{equation}
with $P_k(\xi)$ satisfying
\begin{equation}\label{kdvburgers2reductioneq2}
\nu
P_{k,\xi}-3\sum_{i=0}^kP_{k-i}P_i-P_{k-1,\xi\xi}+a\delta_{k,0}\xi+A_k=0,\
(k=0,\ 1,\ \cdots)
\end{equation}
where $P_{-1}=0$ and $A_k$ are integral constants.

When $k=0$ and $a\equiv-b$, we get the Airy function solution to Eq.
(\ref{kdvburgers2reductioneq2})
\begin{equation}
P_0=\frac{(3b\nu)^\frac{1}{3}[C_1\mathrm{Ai}(1,3^{\frac{1}{3}}(b\nu)^{-\frac{2}{3}}(A_0-b\xi))+
\mathrm{Bi}(1,3^{\frac{1}{3}}(b\nu)^{-\frac{2}{3}}(A_0-b\xi))]}{3[C_1\mathrm{Ai}(
3^{\frac{1}{3}}(b\nu)^{-\frac{2}{3}}(A_0-b\xi))+\mathrm{Bi}(3^{\frac{1}{3}}(b\nu)^{-\frac{2}{3}}(A_0-b\xi))]},
\end{equation}
where $C_1$ is an arbitrary constant. The Airy wave functions ${\rm
Ai}(z)$ and ${\rm Bi}(z)$ are linearly independent solutions for
$y(z)$ in the equation $y''(z)-zy(z)=0$. ${\rm Ai}(n,z)$ and ${\rm
Bi}(n,z)$ are the $n$th derivatives of ${\rm Ai}(z)$ and ${\rm
Bi}(z)$ evaluated at $z$, respectively.

The hyperbolic tangent function solution of traveling wave form
\begin{equation}\label{tanhsol}
P_0=-\frac{\sqrt{3}}{3}p\tanh[\frac{\sqrt{3}}{\nu}p(\xi+d)],
\end{equation}
with $d$ an arbitrary constant, can be obtained from $a=0$ and
$p=\sqrt{A_0}$.

When $k>0$, an equivalent form of Eq.
(\ref{kdvburgers2reductioneq2}) is
$$
\nu P_{k,\xi}-6P_0P_k=g_k(\xi),\ (k=1,\ 2,\ \cdots)\eqno(36')
$$
where
\begin{displaymath}
g_k(\xi)=3\sum_{i=1}^{k-1}P_{k-i}P_i+P_{k-1,\xi\xi}-A_k.
\end{displaymath}
From Eq. $(36')$, it is easily seen that the $k$th similarity
reduction equation in Eq. (\ref{kdvburgers2reductioneq2}) is a first
order linear ordinary differential equation of $P_k$. Furthermore,
all equations in Eq. $(36')$ can be solved step by step. The results
read
\begin{equation}
P_k=\exp(\frac{6}{\nu}\int P_0{\rm d}\xi)\left[\frac{1}{\nu}\int
g_k\exp(-\frac{6}{\nu}\int P_0{\rm d}\xi){\rm d}\xi+B_k\right],\
(k=1,\ 2,\ \cdots)
\end{equation}
where $B_k$ are arbitrary integral constants.

\section{Conclusion and discussion}
In summary, by applying the approximate symmetry reduction approach
to (1+1)-dimensional KdV-Burgers equation under the condition of
weak dispersion and weak dissipation, we have found that the
similarity reduction solutions and similarity reduction equations of
different orders are coincident in their forms. Therefore, we
summarize the series reduction solutions and general formulae for
the similarity equations.

For weak dissipation case, zero-order similarity solutions are
equivalent to Painlev\'e II, Painlev\'e I type and Jacobi elliptic
function solutions. For weak dispersion case, zero-order similarity
solutions are in the form of Kummer function, Airy function and
hyperbolic tangent function solutions.

$k$-order similarity reduction equations are linear ordinary
differential equations with respect to $P_k(\xi)$. Especially, for
the period solutions (expressed by Jacobi elliptic functions) with
solitary waves as a special case under weak dissipation, Airy
function and hyperbolic tangent function solutions under weak
dispersion, all the higher order similarity solutions can be solved
simply by direct integration.

The convergence of infinite series solutions remains as a problem
and expects further study. The approximate symmetry reduction
approach can be used to search for similar results of other
perturbed nonlinear differential equations and it is worthwhile to
summarize a general principle for the perturbed nonlinear
differential equations holding analogous results.

\begin{acknowledgments}
The work was supported by the National Natural Science Foundations
of China (Nos. 10735030, 10475055, 10675065 and 90503006), National
Basic Research Program of China (973 Program 2007CB814800) and
PCSIRT (IRT0734), the Research Fund of Postdoctoral of China (No.
20070410727) and Specialized Research Fund for the Doctoral Program
of Higher Education (No. 20070248120).
\end{acknowledgments}

\end{document}